\documentclass{article}
\usepackage[T1]{fontenc} 
\usepackage[utf8]{inputenc} 
\usepackage{myismir,amsmath,cite,url} 
\usepackage{graphicx}
\usepackage{color}
\usepackage{float}

\title{Towards Cross-Cultural Analysis using Music Information Dynamics}

\threeauthors
  {Shlomo Dubnov} {UCSD \\ {\tt sdubnov@ucsd.edu}}
  {Kevin Huang} {UCSD \\ {\tt kyh022@ucsd.edu}}
  {Cheng\-i Wang} {Smule \\ {\tt cheng-i.wang@smule.com}}

\date{}

\begin{document}

\maketitle

\begin{abstract}
A music piece is both comprehended hierarchically, from sonic events to melodies, and sequentially, in the form of repetition and variation. Music from different cultures establish different aesthetics by having different style conventions on these two aspects. We propose a framework that could be used to quantitatively compare music from different cultures by looking at these two aspects.

The framework is based on an Music Information Dynamics model, a Variable Markov Oracle (VMO), and is extended with a variational representation learning of audio. A variational autoencoder (VAE) is trained to map audio fragments into a latent representation. The latent representation is fed into a VMO. The VMO then learns a clustering of the latent representation via a threshold that maximizes the information rate of the quantized latent representation sequence. This threshold effectively controls the sensibility of the predictive step to acoustic changes, which determines the framework's ability to track repetitions on longer time scales. This approach allows characterization of the overall information contents of a musical signal at each level of acoustic sensibility.

Our findings under this framework show that sensibility to subtle acoustic changes is higher for East-Asian musical traditions, while the Western works exhibit longer motivic structures at higher thresholds of differences in the latent space. This suggests that a profile of information contents, analyzed as a function of the level of acoustic detail can serve as a possible cultural characteristic.

\end{abstract}
\section{Introduction}\label{sec:introduction}

The music information retrieval (MIR) community in recent years is encouraging more MIR tasks to be applied to non-western music \cite{Lidy2010,Serra2011,Serra2017}, with comparative analysis between music from different cultures or cross-cultural music analysis becoming a rapid developing research area. In this paper we prose a framework extending from a previous work done in \cite{Dubnov2016} on comparing music recordings from different cultures using a computational framework of Music Information Dynamics. The proposed framework aims at providing quantitative measurements of musical structure based on information theoretical considerations. In this paper we show how this methods could be applied to show stylistic differences between music from different cultures. 

In \cite{Panteli2017}, hand-crafted audio features and traditional machine learning techniques are used to construct a similarity measurement to find outliers among music examples from various cultural origins and analyze their characteristics. In \cite{Panteli2017a}, hand-crafted vocal features are devised to study different singing styles from different cultures, and in \cite{Vidwans2020}, the same work is extended to classify concert non-western music recordings. Various non western or folk music corpus are evaluated and compared in \cite{Savage2018}. To the extent of the authors' knowledge, most of the previous works on cross-cultural comparative music analysis used hand-crafted audio features and traditional machine learning techniques, and none of the works attempted to model how temporality of music is realized differently in different cultures. It is a well established universal observation that music is an temporal art form \cite{Stambaugh1964} that is constructed via sequential relationships, repetitions and variations of musical elements \cite{Kivy1993,Campbell2010}, realized differently in different cultures. 

The previous paragraph brings us to the core motivation of the work proposed in this paper, that is, a music piece is simultaneously comprehended hierarchically and sequentially. It is hierarchical in the sense that, a basic ``sound unit'' could be identified from timbral variations to microtonal expressions to pitched events in a rigid tuning system. The comprehension is sequential since it is the temporal relationships of these sound units that forms larger/longer musical concepts such as phrases, progressions, repetitions and variations. Given the above discussion, we argue that, to compare music from different cultures quantitatively, a proposed framework must takes both hierarchical and sequential relationships into considerations, to reflect the style and aesthetic differences among them.  

We simplify the concurrences of hierarchical and sequential comprehension on a music piece by dividing the process into two steps. In the first step, a compressed representation is learned by a variational autoencoder (VAE) \cite{Kingma2019} from the music signal to best reconstruct itself. Given an input in time-frequency format, a VAE assumes little about either the timbral or tonal structure of the input, which suits the purpose in our case to universally model the ``sound unit'' with little presumption beforehand. Then in the second step, to capture temporal relationships, we adopted the Variable Markov Oracle (VMO) \cite{Wang2015a} that is capable of discovering repeated patterns from the input signal. The VMO is motivated by Music Information Dynamics that maximizes an approximated mutual information measurement that optimizes the balance between repetitions and variations embedded in the input signal. 

The paper is structured as follows. In section \ref{sec:info_dyna} we first describe the Music Information Dynamics and the VMO algorithm. The framework including the VAE is elaborated in section \ref{sec:method}. Experiments on music examples from different cultures are presented in section \ref{sec:exp}. Discussions and conclusions are in section \ref{sec:conclusion}.

\section{Music information dynamics}\label{sec:info_dyna}

Assuming the music signal $X=x[n]$ is encoded into a sequence of ``sound units'' $Z=z[n]$, with $n$ denotes discrete time step $n$. We would like to algorithmically discover the sequential structure of $Z$, and be able to present the structures quantitatively. Music Information Dynamics (MID)\cite{Abdallah2009,Dubnov2011a,Pearce2018} provides a theoretical framework utilizes mutual information between past and present observations to model the predictability of the signal. The advantage of adopting MID is that it optimizes or calculates an information theoretic measurements on the input sequence $Z$ and is agnostic of specific sequence related applications, such as motifs discovery or structure segmentation. MID was shown to be important for understanding human perception of music in terms of anticipation and predictability \cite{Abdallah2009,Dubnov2011a}. 
    
An efficient formal method for studying MID for sequence $Z[n]$ is the Information Rate (IR) that considers the relation between the present measurement $Z=z[n]$ and it's past $\overleftarrow{Z}=z[1],z[2],\dots,z[n],\dots,z[N]$, formally defined as the maximum of mutual information over different quantized level of the sequence $S=Q(Z)$
    
\begin{align}
    IR(Z) & = \max_{Q: S=Q(Z)} I(Q(Z),Q\overleftarrow{(Z)}) \\
    & = H(S)-H(S|\overleftarrow{S})
\end{align}
   
According to this measure, the maximal value of IR is obtained when the difference between the uncertainty of $H(S)$ and predictability $H(S|\overleftarrow{S})$ is at its greatest, meaning that there is a balance between variation and predictability. Quantization $Q(Z)$ is needed due to the need to detect inexact repetitions in the sequence $Z$, which in turn signifies the allowed level of similarities between observations in $Z$, or the amount of ``sound unit'' details that is significant when comparing the present to the past. 

One of the realization of MID as an algorithm is the Variable Markov Oracle (VMO) \cite{Wang2015a}. Variable Markov Oracle (VMO) is a generative machine learning method that combines lossy compression with the Factor Oracle (FO) string matching algorithm\cite{Allauzen1999}. 
VMO is capable of finding embedded linkages between samples along the multivariate time series and enables tracking and comparison between time-series using a Viterbi-like dynamic programming algorithm. In order to operate on such multivariate time series data, \emph{VMO} symbolizes a signal $Z$ sampled at time $n$ into a symbolic sequence $S = s[1],s[2],\dots,s[n],\dots,s[N]$, with $M$ states and with observation frame $z[n]$ labeled by $s[n]$ whose value is one of the symbols in a finite sized alphabet $\Sigma$. The labels are formed by following suffix links along the states in the suffix tree structure constructed by the VMO algorithm. 

The essential step for the construction of VMO is finding a threshold value, $\theta$ that partitions the space of features into categories. The threshold $\theta$ is used to determine if the incoming $z[n]$ is similar to one of the frames following the suffix link started at $n-1$. VMO assigns two frames, $z[i]$ and $z[j]$, the same label $s[i]=s[j] \in \Sigma$ if $||z[i]-z[j]||\le\theta$. In extreme cases, setting $\theta$ too low leads to VMO assigning different labels to every frame in $Z$ and setting $\theta$ too high leads to VMO assigning the same label to every frame in $Z$. As a result, both extreme cases are incapable of capturing any temporal structures (repeated suffixes) of the time series. To find the optimal threshold $\theta$, the aforementioned IR is approximated by a deterministic notion of a compression algorithm $C(\cdot)$ that is used as a computational measure of time series complexity\cite{Lefebvre2002}. This allows estimation of the mutual information by replacing the the entropy term $H(\cdot)$ in with compression gain for the best quantization, searching over possible values of the  threshold $\theta$ (Eq. \ref{eq:ir}).   

\begin{equation}\label{eq:ir}
IR(Z)= \max_{\theta, s[n] \in \Sigma_{\theta}} [C(s[n])-C(s[n]| \overleftarrow{S})].
\end{equation}

Due to space consideration, we leave out the VMO construction and IR optimization algorithms and refer the readers to \cite{Dubnov2011,Wang2015a}. It should be noted that the alphabet out of the quantization is constructed dynamically, as new labels can be added when an input sample cannot be assigned to one of the existing clusters of samples already labeled by existing labels. We will denote the resulting alphabet for a given $\theta$ as $\Sigma_{\theta}$.


\section{Methodology}\label{sec:method}
In this section we describe the comparative framework by introducing each component. Given a music recording, a time-frequency representation $X$ with $N$ frames is first extracted and fed into the VAE to learn a compressed representation $Z$ for each frame. Then $Z$ is quantized and indexed by the VMO algorithm given a range of thresholds. Two quantitative results are obtained via VMO. The first is an IR profile by plotting the IR values as a function of the threshold. The second is motif discovery results obtained by the optimal threshold $\theta$ that returns the highest IR value. Comparisons between recordings could then be made by examine these two outputs. 

\subsection{Representation Learning}\label{sec:vae}

In this work, as discussed above, we first need a representation learning step to extract the ``sound units'' from the audio signal. The high dimensional and continuous nature of audio signals creates a challenge for finding an efficient latent representation for music. Several works have utilized the VAE models for finding disentangled representations in audio \cite{Luo2019} and modeling audio containing speech \cite{Hsu2017}. In this work we use the variational autoencoder (VAE) \cite{Kingma2019} to map standard time-frequency representations into latent variables $Z$ that captures the intrinsic sonic/tonal structures of the sound.  A normal autoencoder is a deep learning structure that contains an encoder and a decoder, each of them comprised of trainable non-linear transformations made by neural networks. During the training process, an input is encoded into a latent vector, also known as an embedding, and an output is then decoded from the same latent vector. The goal of training an autoencoder is to make the input and output as similar as possible, while trying to obtain an embedding that has some desired properties, such as dimensionality reduction, sparseness and so on. Broadly speaking, an autoencoders is an unsupervised learning algorithm, since the goal is to reconstruct itself by learning a compressed representation.

A VAE explicitly constrains the latent variables $Z$ so that they are distributed according to a prior $p(z)$. The input $X$ and latent code $Z$ can then be seen as random variables $Z\sim p(Z), X\sim p(X|Z)$. The VAE consists of an encoder probability $q_\lambda(Z|X)$, which approximates the posterior probability $p(Z|X)$, and a decoder probability $p_\theta(X|Z)$, which parameterizes the likelihood $p(X|Z)$. In practice, the approximate posterior and likelihood distributions are parameterized by the weights the neural networks. Posterior inference is done by minimizing the KL divergence between the encoder and the true posterior. It can be proved that this optimization problem is the same as maximizing the evidence lower bound (ELBO):

\begin{equation}
\begin{split}
ELBO &= E[\log p_\theta(X|Z)]-KL(q_\lambda(Z|X)||p(Z)) \\
&\leq \log p(X)
\end{split}
\end{equation}

\subsection{Temporal Structure Discovery}

After a sequence of learned representation $Z$ is obtained from the VAE, the VMO is applied on $Z$ to discover the embedded repetitive structure. As discussed above, a VMO captures the optimal temporal structure of a signal by adapting the threshold $\theta$ to distinguish between frames that maximizes the IR value. This tuning can be considered as a way to extract the most balanced repetitions and variations from the signal given a trade-off of tolerance to imprecise matching between repeating signal segment. To put it differently, the IR value at each threshold represents that total IR at that level of detail. If then the IR values are plotted as a function of thresholds, we obtain an ``IR profile'' that shows how sensitive the IR of the signal is to the level of quantization. Motifs or repeated patterns and their statistics from the quantized sequence $S$ from $Z$ obtained via VMO is also used to compare among different music recordings. The motif discovery algorithm based on VMO is documented in \cite{Wang2015a}. In short, the VMO motif discovery algorithm is a repeated sub-string finding algorithm given a suffix tree on the original string.  

\section{Experiments}\label{sec:exp}

In our experiments, the audio recording is sampled at 22050Hz and then a time-frequency representation $X$ is obtained by an STFT with a window length of 2048 and a hop length of 256 frames. The spectrogram is then split into short and non-overlapping frames with frame size of 24, resulting in an input representation with dimensions $1024\times 24$. The encoder and decoder of our VAE have a symmetrical structure consisting of convolutional layers and linear layers. Full details for the architecture parameters are available in Appendix \ref{appendix:a}. The reconstruction loss function is the mean-squared error between the input and reconstructed spectrograms, and the whole VAE training loss combines the reconstruction loss and the KL divergence loss. For the experiment, we also extracted hand-crafted features such as MFCCs and Chroma features with the same STFT parameters. The MFCCs are calculated by 128 bands and we take the first 19 MFCCs omitting the 1st coefficient. Both Chroma features and MFCCs are mean-aggregated every 24 frames to have the same time resolution as the input spectrogram frames to the VAE. To train a VAE, we use a batch size of 50 with total 250 epochs optimized by ADAM \cite{Kingma2015}. 

\subsection{Choice of repertoire}
We performed comparative analysis on four musical works for the flute, two from Western cultures and two from East-Asian traditions. The Western works considered here are Bach, Partita in A Minor, BWV 1013 and Telemann: Fantasia No. 3 in B minor. The East-Asian works are one Shakuhachi traditional Japanese piece and one Nanguan classical Chinese music performed on a Dongxiao flute. 

The duration of the pieces were between 4-5 minutes, except for the Bach Partita that was around 9 minutes in duration. This particular choice of repertoire is intentionally limited to instruments of similar timbre and monophonic music material, while at the same time still allowing for different levels of expressive inflections that are typical to different cultures. In order to test for differences between different types of flutes, we compared two performances of Teleman's Fantasia, one performed on a modern flute (traverse metal flute) and the other performed on a recorder (wooden beak flute). The recorder, shakuhachi and dongxiao flutes are end blown, but unlike the recorder, where the player blows into an internal whistle like duct, the Shakuhachi and Dongxiao players blow across the flute, which results in different levels of pitch and breath control.

\begin{figure*}[t]
    \centering
    \includegraphics[width=0.8\linewidth]{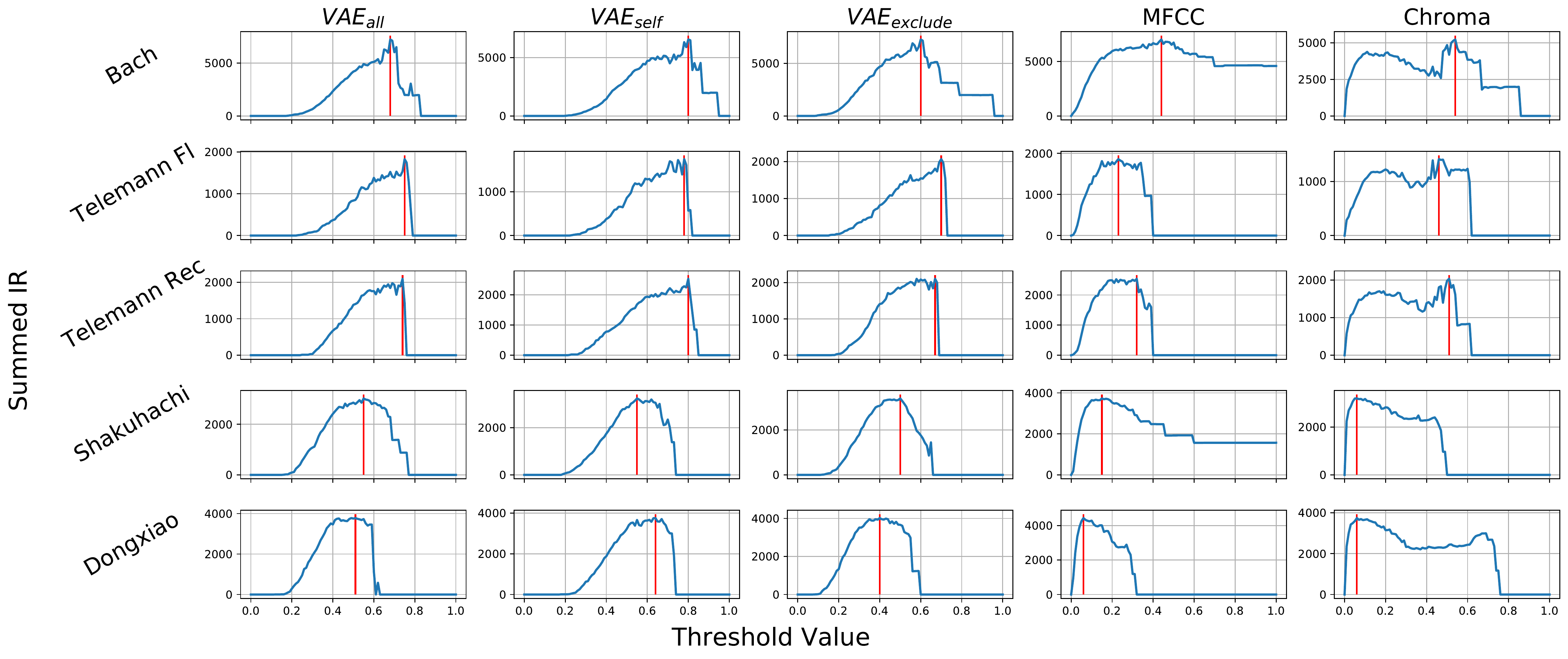}
    \caption{Information Rate profiles for the three VAE methods, MFCCs, and Chroma features, for all five flute pieces}
    \label{FluteIR}
\end{figure*}

\subsection{Analysis Method}
As mentioned earlier, the analysis was done in two separate steps, the first being a feature extraction that was done by representation learning using VAE, followed by a second step of temporal structure analysis using VMO. In order to get a better insight into the role of learned representations in musical information dynamic (MID), we used Mel-frequency cepstral coefficients (MFCC) and Chroma features\footnote{By librosa https://github.com/librosa/librosa} as baselines for comparing MID analysis between hand-crafted features and learned representations. Musically speaking, MFCC and Chroma correspond to timbral and tonal properties of the music recordings, as the MFCCs capture the broad spectral shape of the timbre of the instrument, while Chroma features wrap the frequency analysis into bins corresponding to the Western twelve tone system. In a sense, these hand-crafted features incorporate prior cultural assumptions as they are designed to capture different aspects of the acoustic signal that might be prevalent in certain musical practices more than in others. Specifically, the Chroma features are designed so as to be invariant to delicate timbral variations by aggregating the spectral features into $12$ bins that correspond to well defined pitches that are predominantly used in Western musical culture. As such, it is natural to consider if a learned representation (on the task of reconstructing itself) can be agnostic to cultural-biases. Furthermore, by training on different sets of the recordings, it is possible to experiment if learning either a piece, a cultural, or an instrument specific representations. Accordingly, in our experiments we compared three different types of VAE representations, one that was trained on all musical examples across cultures (ALL-method), one that was trained on the same musical piece that VMO analyzed (SELF-method), and one that was trained on all of the examples excluding the specific piece that was analyzed by the VMO (EXCLUDE-method). 

\subsection{IR Profiles}\label{subsec:ir}
For each of the five music pieces (Bach, Telemann flute, Telemann recorder, Shakuhachi and Dongxiao), we apply the three (ALL, SELF, and EXCLUDE) methods of training VAE, and extract MFCCs and Chroma features on it. Each of these conditions resulted in a sequence $Z$ that we feed it into the VMO algorithm, and calculate a IR value for a range of threshold to obtain an IR profile. Cosine distance is used as the distance function in $||z[i]-z[j]||\le\theta$ for each $\theta$ value. 

The result IR profiles are plotted in Figure \ref{FluteIR}. In each subplot, IR value is on the vertical axis and thresholds are on the horizontal axis. It is obvious that Western pieces tend to have a large peak at higher thresholds, while East-Asian examples tend to have higher IR at lower threshold values. Implicating that the East-Asian examples utilize more subtle timbral variation to create phrases and expressions, then build repetition and structures on top of them. For Western examples, this observation corresponds to the fact that these pieces are more tonal driven. 

\subsection{Motif Discovery}\label{subsec:motif}
\begin{figure}[]
    \centering
    \includegraphics[width=0.9\columnwidth]{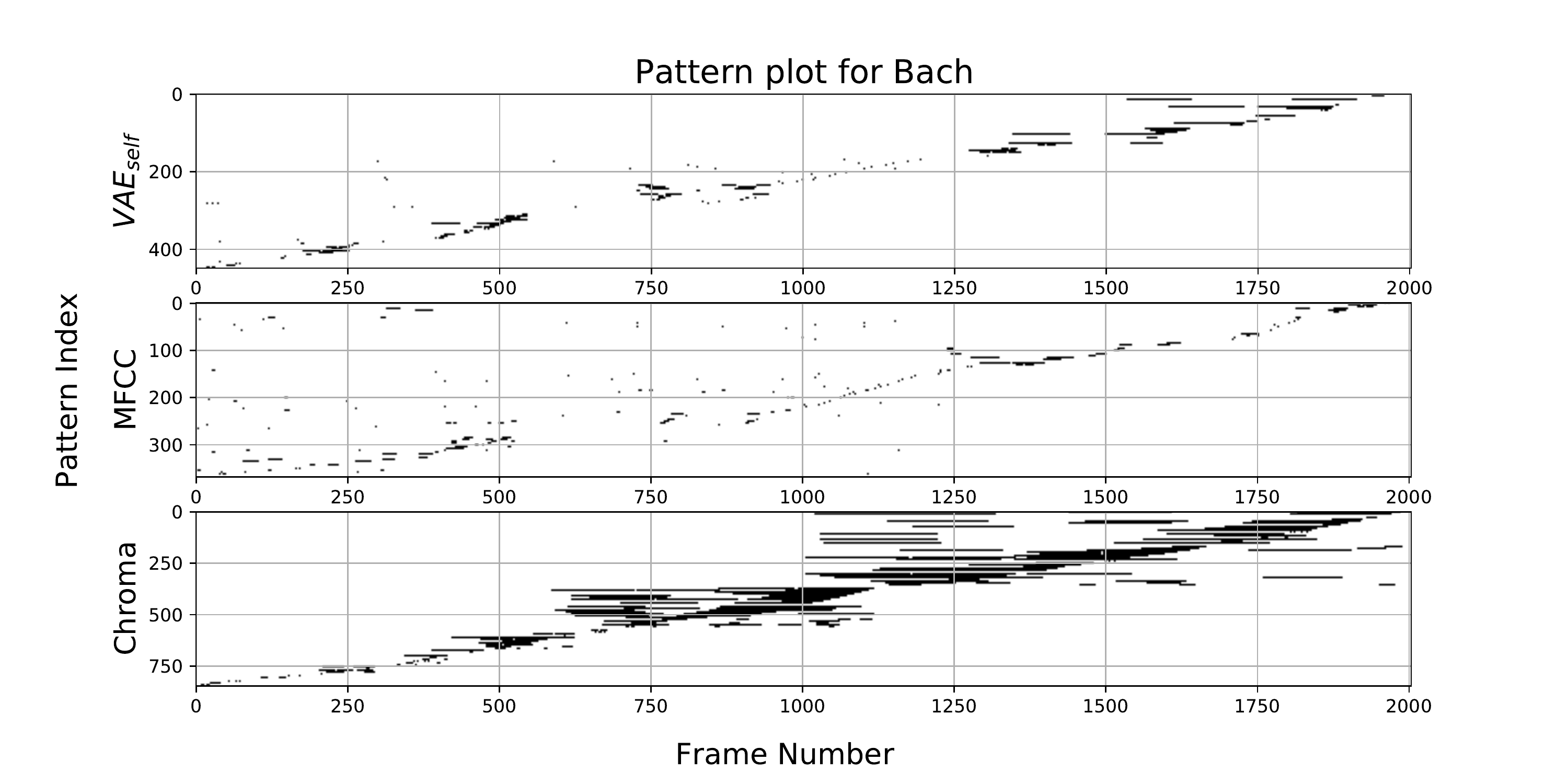}
    \includegraphics[width=0.9\columnwidth]{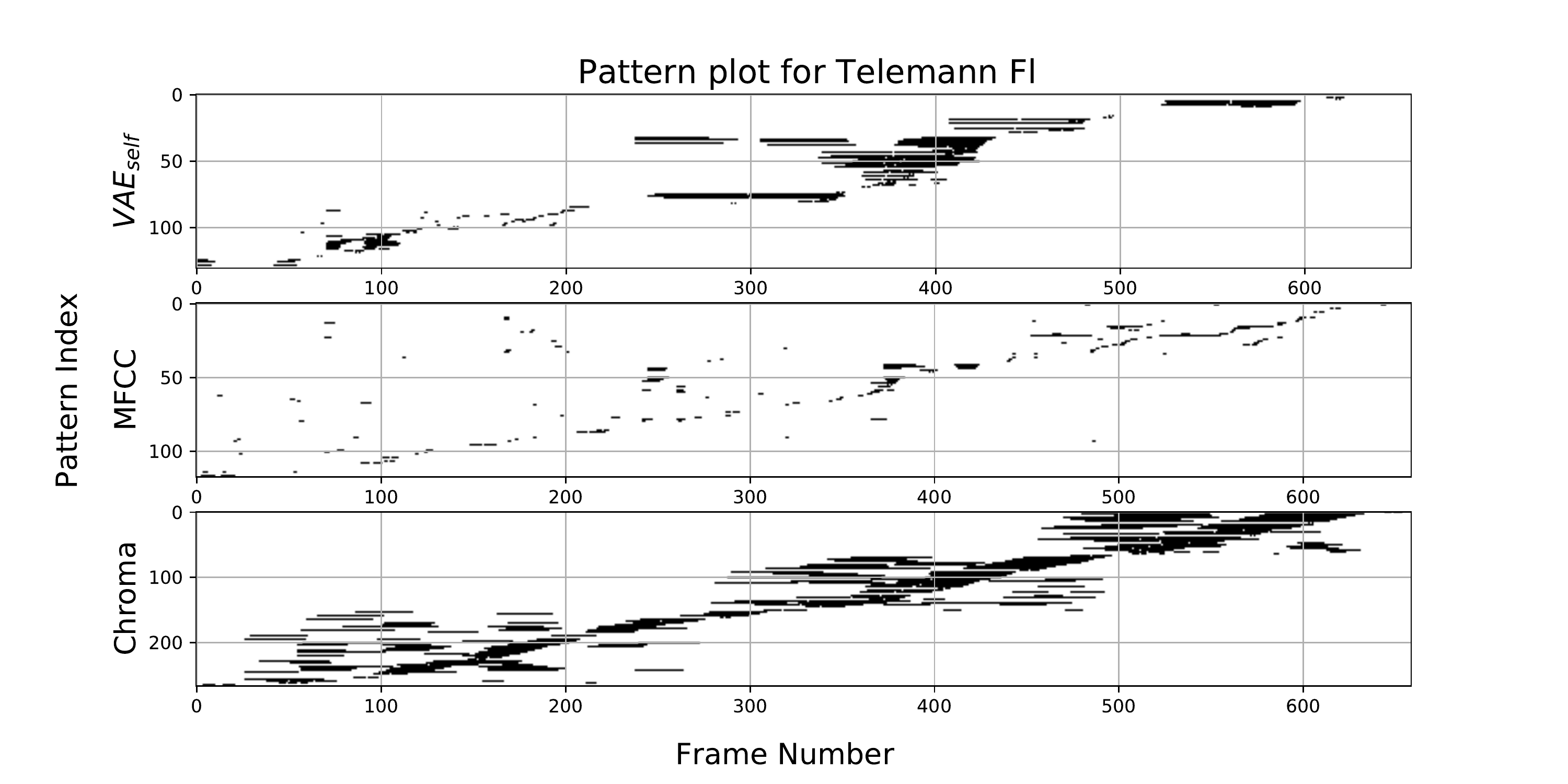}

    \includegraphics[width=0.9\columnwidth]{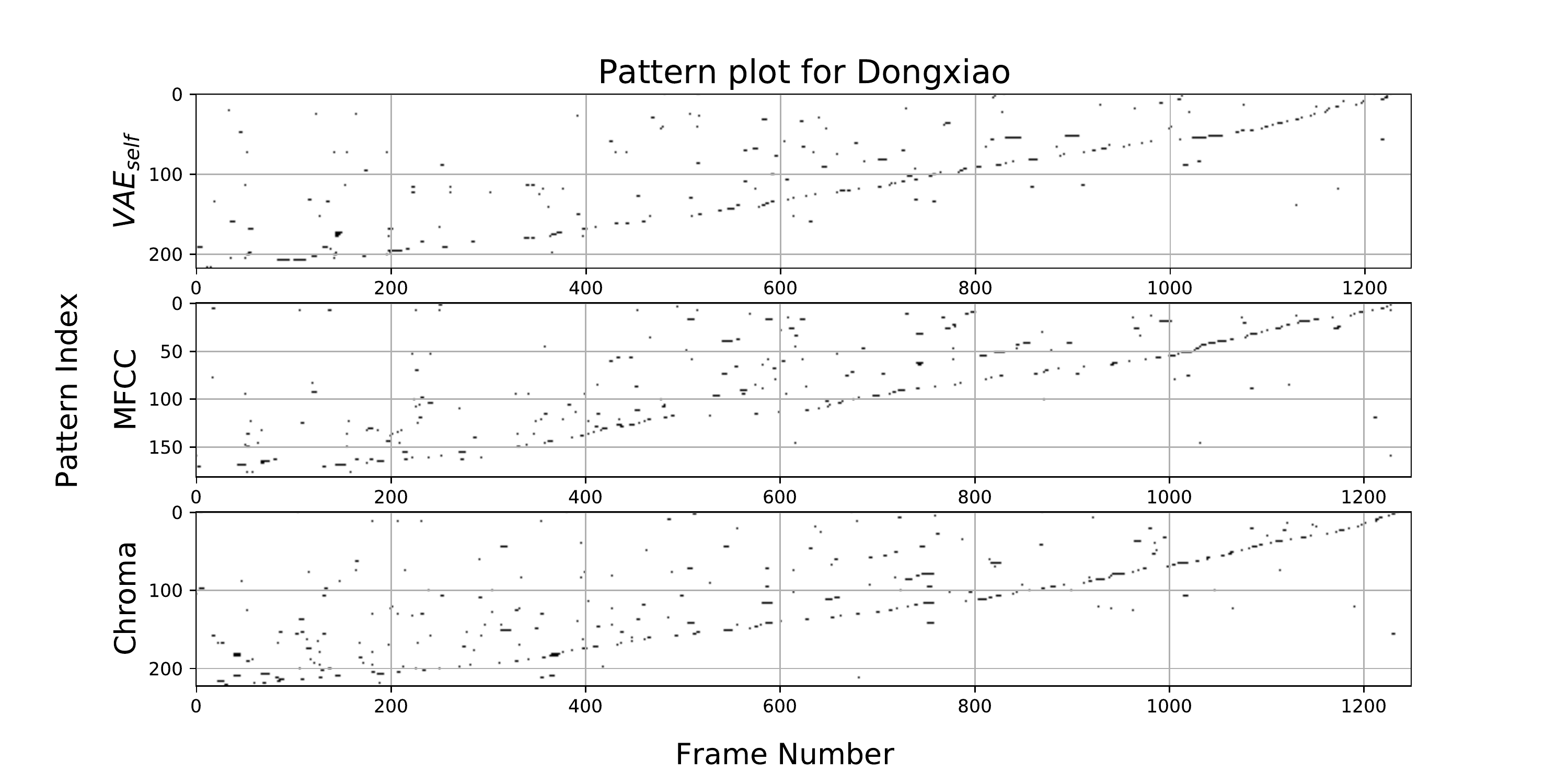}
    \caption{Motifs found in three flute pieces using the best VMO for $\text{VAE}_{self}$, Chroma features, and MFCCs.}
    \label{MotifsFound}
\end{figure}

For each of the five pieces, we also apply the VMO motif discovery algorithm to obtain a list of repeated sub-sequences. In this case, we only return the motifs found via the $\theta$ that maximized the IR values(The red vertical lines in Figure \ref{FluteIR}). For space consideration, we choose Bach, Telemann Flute, Dongxiao and only the SELF-method to demonstrate the results. The results are depicted in Figure \ref{MotifsFound}. In each subplot, the horizontal axis is the time step and the vertical indices refers to the found motifs. Each horizontal black line represent an occurrence of a single motif, and the lines on the same vertical positions are repetitions of one another. The length of lines represents the length of the motifs. The ordering of the motifs is from the end of the piece at the top, to the beginning at the bottom. The ordering was done for plotting conveniences. By looking at the plots it is also noticeable that both Western examples have longer motifs and are mostly in the tonal space, while the dongxiao piece has much shorter motifs and the similar amount of motifs can be found either via timbre or pitch information. 

\subsection{Representations}
By looking at the results in section \ref{subsec:ir} and Figure \ref{FluteIR}, the observation is that the VAE feature learning is invariant to the three training methods experimented. The results from SELF, EXCLUDE and ALL are all relatively consistent with each other. In a way this result is not surprising given that all the pieces are flute pieces. On the differences between hand-crafted features and learned representations, by looking at the results from section \ref{subsec:ir} and \ref{subsec:motif}, we can verify that although the results from learned representations are less distinguishable compared to hand-crafted features (IR peaks between pieces are closer), but they are still consistent with the results from hand-crafted features. Also by looking at the motif discovery results, we can observe that although the VAE is not trained toward any indication of temporal relationships between frames in $X$, we can still find meaningful motifs on the resulted $Z$ sequence, implicating VAE's ability as a universal encoder. The advantage of using learned representations is still that it posses less algorithmic biases.    



\begin{figure}
\begin{center}
\includegraphics[width=0.99\columnwidth]{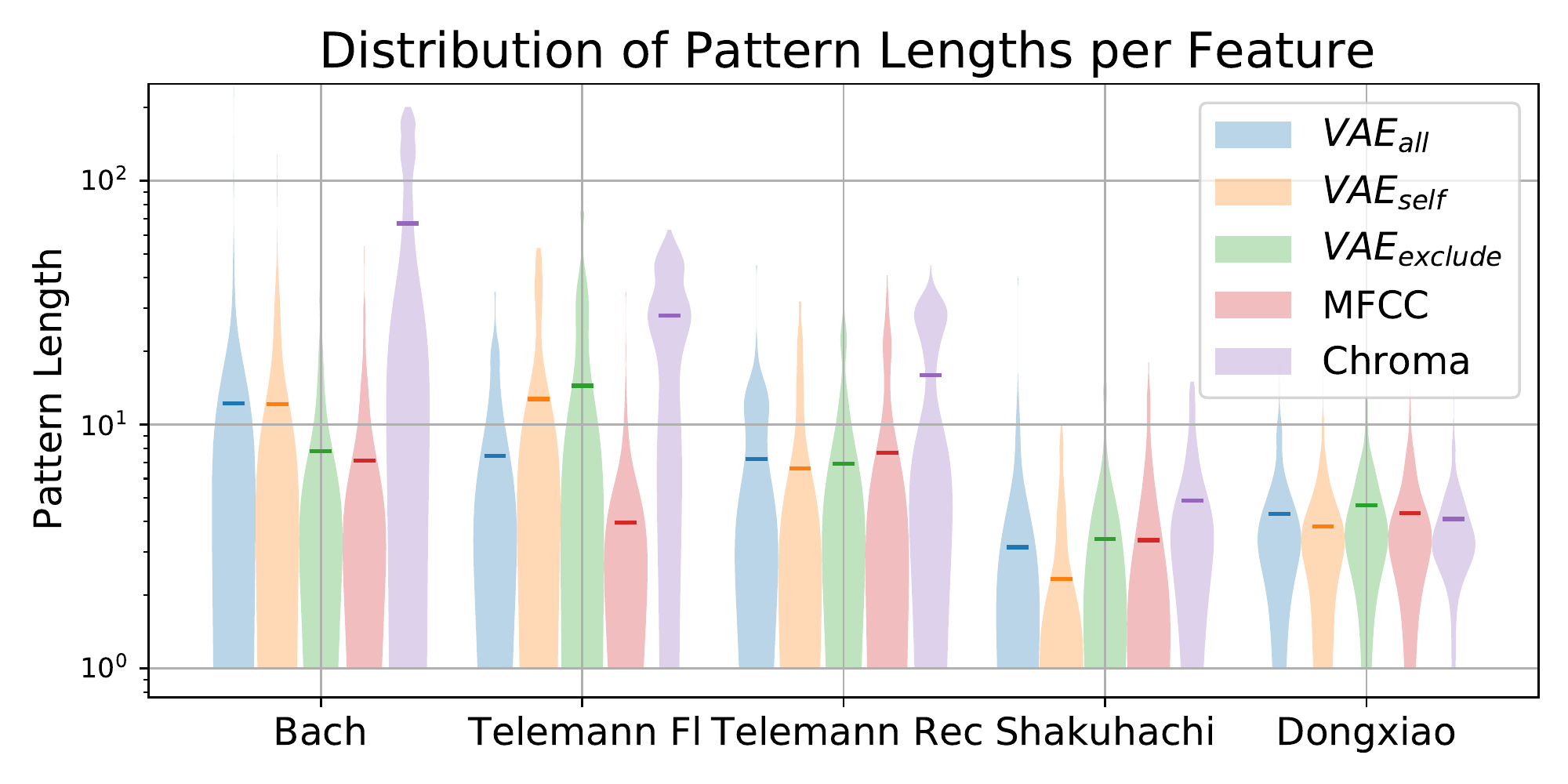}
\caption{Mean and distribution of the motif lengths for the VAE, Chroma, and MFCC features}
 \label{MotifLengthsStatistics}
\end{center}
\end{figure}

\begin{table}
    \centering
    \begin{tabular}{c|c c}
        Feature & F-Value & p-value \\\hline
        $VAE_{all}$ & 20.859 & $5.499\times10^{-6}$\\
        $VAE_{self}$ & 58.169 & $5.439\times10^{-14}$\\
        $VAE_{exclude}$ & 49.190 & $4.354\times10^{-12}$\\
        MFCC & 31.082 & $4.150\times10^{-8}$\\
        Chroma & 316.287 & $4.179\times10^{-63}$
    \end{tabular}
    \caption{ANOVA test results comparing the average sequence length in Eastern and Western music}
    \label{tab:anova_motif}
\end{table}

\section{Discussion and Conclusion}\label{sec:conclusion}
In order to obtain more quantitative estimate of the differences in motif structure among the different pieces, the mean and overall distribution of motifs found at the last step are summarized in Figure \ref{MotifLengthsStatistics}. Also the ANOVA tests are summarized in Table \ref{tab:anova_motif}. We found there to be a significant difference in the mean motif length between our selections of East-Asian and Western pieces. Further statistical analysis can be found in Appendix \ref{appendix:b}.These results indicate that for Western works the VMO algorithm tends to find more structure at the higher threshold levels, thus favoring coarser structure or units that quantize larger differences between the element in $z$ under one label. This results in significantly longer motifs for all Western pieces. It is interesting to observe that Telemann piece performed with a recorder has a somewhat larger peak at a lower threshold. We interpret this as an indication that there is increased structure at the finer sonic details, that might be related to the nature of the recorder instrument. 
It should be noted that the method of motif discovery is based on a single VMO at the optimal threshold. It is possible that multiple peaks of comparable level in terms of their IR value would appear for a particular piece. In such a case, it is possible to interpret such situation as indication that significant musical structure exists at different threshold levels. In such case it is not clear which of the thresholds is preferable for performing the motif analysis. In other words, performing syntagmatic analysis (motif finding) depends on the choice of the aesthetic criteria (IR optimum). Our experience shows that lower thresholds that correspond to finer acoustic differences tend to result in shorter motifs. For instance, analysis of the Telemann piece performed with the recorder at a low threshold value that corresponds to the secondary peak results in significantly shorter motifs that is comparable to the lengths found in the Shakuhachi and Dongxiao recordings. 

The next steps are first to examine the different learned $Z$ under different configurations to understand if we can find meaningful insight on the differences between music from different cultures on the sound level, and how much a learned representation (such as the one learned by a VAE) could be generalized. Then we would like to explore the possibility of combining the two steps, representation learning and temporal structure discovery, into one holistic model, to closer mimic the concurrences of hierarchical and sequential comprehension of a music piece. In parallel we will continue to experiment on different families of instruments and music from different cultures. 

It is also important to realize that the retrieved motifs using VMO, which is a technique fundamental to the analysis presented in this paper, might differ from other motif analysis methods, such as those in MIR challenges for motif analysis. In \cite{Wang2015} the VMO motif analysis results were compared to selected musical motif discovery systems and showed SOTA performance at the time of the writing (2015). It is possible that other motif techniques would render different results that could be then compared based on their musical significance. To the best of our knowledge, there is little evidence of analyzing non-Western music of the type we discuss in this paper in terms of generalized motifs. We consider any sequence of timbral inflections, as captured by their embeddings and clustering, as motifs. These non-conventional motif maybe not "musical" motifs in the traditional sense, but rather "statistical" motifs as they represent sequences of audio features that have salient repetition statistics. Accordingly, the use of the term "motif" should be read in a somewhat ambiguous way denoting a structural musical element over time, rather then engaging in musicological analysis of the correctness/relevance/meaningfulness of the discovered motifs.

The code with complete graphs can be found in the code repository\footnote{https://github.com/Origamijr/Cross-Cultural-Information-Dynamics}.

\subsection{Cultural Disclaimer}

The analysis of the four pieces presented in this paper does not shed light upon the vastness of Western music and the larger vastness of East Asian music, so the results presented here should not be interpreted as characteristics of the whole music cultures. To establish more solid knowledge of structural and timbral aspects of even just Japanese flute repertory, repertoires for folk shinobue or theatrical nohkan should be included. Equally, nanguan might be more representative of the regional style from South-East China, which is very different from flute music from Nothern China or Korea. This could be compared with the very music for Daegeum that has strong characteristics in terms of timbre and structure. Equally, historically speaking, Telemann and Bach are only minimal representatives of central European late Baroque music, which is very differnt  from romantic, or avant-garde music for flute, as well as other folk flute traditions in Europe. 
Therefore, the results of these experiments should only be considered as an initial pilot and a limited use case for a method which could be applied to way larger corpora in order to shed light upon such cultural differences.

\bibliography{ISMIR_2021_cross_cultural}

\begin{thebibliography}{10}
\providecommand{\url}[1]{#1}
\csname url@samestyle\endcsname
\providecommand{\newblock}{\relax}
\providecommand{\bibinfo}[2]{#2}
\providecommand{\BIBentrySTDinterwordspacing}{\spaceskip=0pt\relax}
\providecommand{\BIBentryALTinterwordstretchfactor}{4}
\providecommand{\BIBentryALTinterwordspacing}{\spaceskip=\fontdimen2\font plus
\BIBentryALTinterwordstretchfactor\fontdimen3\font minus
  \fontdimen4\font\relax}
\providecommand{\BIBforeignlanguage}[2]{{%
\expandafter\ifx\csname l@#1\endcsname\relax
\typeout{** WARNING: IEEEtran.bst: No hyphenation pattern has been}%
\typeout{** loaded for the language `#1'. Using the pattern for}%
\typeout{** the default language instead.}%
\else
\language=\csname l@#1\endcsname
\fi
#2}}
\providecommand{\BIBdecl}{\relax}
\BIBdecl

\bibitem{Lidy2010}
T.~Lidy, C.~N. Silla~Jr, O.~Cornelis, F.~Gouyon, A.~Rauber, C.~A. Kaestner, and
  A.~L. Koerich, ``On the suitability of state-of-the-art music information
  retrieval methods for analyzing, categorizing and accessing non-western and
  ethnic music collections,'' \emph{Signal Processing}, vol.~90, no.~4, pp.
  1032--1048, 2010.

\bibitem{Serra2011}
X.~Serra, ``A multicultural approach in music information research,'' in
  \emph{Klapuri A, Leider C, editors. ISMIR 2011: Proceedings of the 12th
  International Society for Music Information Retrieval Conference; 2011
  October 24-28; Miami, Florida (USA). Miami: University of Miami; 2011.}\hskip
  1em plus 0.5em minus 0.4em\relax International Society for Music Information
  Retrieval (ISMIR), 2011.

\bibitem{Serra2017}
------, ``The computational study of a musical culture through its digital
  traces,'' \emph{Acta Musicologica}, vol.~89, no.~1, pp. 24--44, 2017.

\bibitem{Dubnov2016}
S.~Dubnov, K.~Burns, and Y.~Kiyoki, \emph{Cross-Cultural Multimedia Computing:
  Semantic and Aesthetic Modeling}.\hskip 1em plus 0.5em minus 0.4em\relax
  Springer, 2016.

\bibitem{Panteli2017}
M.~Panteli, E.~Benetos, and S.~Dixon, ``A computational study on outliers in
  world music,'' \emph{Plos one}, vol.~12, no.~12, p. e0189399, 2017.

\bibitem{Panteli2017a}
M.~Panteli, R.~Bittner, J.~P. Bello, and S.~Dixon, ``Towards the
  characterization of singing styles in world music,'' in \emph{2017 IEEE
  International Conference on Acoustics, Speech and Signal Processing
  (ICASSP)}.\hskip 1em plus 0.5em minus 0.4em\relax IEEE, 2017, pp. 636--640.

\bibitem{Vidwans2020}
A.~Vidwans, P.~Verma, and P.~Rao, ``Classifying cultural music using melodic
  features,'' in \emph{2020 International Conference on Signal Processing and
  Communications (SPCOM)}.\hskip 1em plus 0.5em minus 0.4em\relax IEEE, 2020,
  pp. 1--5.

\bibitem{Savage2018}
P.~E. Savage, ``An overview of cross-cultural music corpus studies,'' 2018.

\bibitem{Stambaugh1964}
J.~Stambaugh, ``Music as a temporal form,'' \emph{The Journal of Philosophy},
  vol.~61, no.~9, pp. 265--280, 1964.

\bibitem{Kivy1993}
P.~Kivy, \emph{The fine art of repetition: Essays in the philosophy of
  music}.\hskip 1em plus 0.5em minus 0.4em\relax Cambridge University Press,
  1993.

\bibitem{Campbell2010}
E.~Campbell, \emph{Boulez, music and philosophy}.\hskip 1em plus 0.5em minus
  0.4em\relax Cambridge university press, 2010, vol.~27.

\bibitem{Kingma2019}
D.~P. Kingma and M.~Welling, ``An introduction to variational autoencoders,''
  \emph{arXiv preprint arXiv:1906.02691}, 2019.

\bibitem{Wang2015a}
C.-i. Wang, J.~Hsu, and S.~Dubnov, ``Music pattern discovery with variable
  markov oracle: A unified approach to symbolic and audio representations,'' in
  \emph{International Society for Music Information Retrieval Conference},
  2015, pp. 176--182.

\bibitem{Abdallah2009}
S.~Abdallah and M.~Plumbley, ``Information dynamics: Patterns of expectation
  and surprise in the perception of music,'' \emph{Connect. Sci}, vol.~21, no.
  2-3, pp. 89--117, Jun. 2009.

\bibitem{Dubnov2011a}
S.~Dubnov, ``Musical information dynamics as models of auditory anticipation,''
  in \emph{Machine Audition: Principles, Algorithms and Systems.}, W.~Wang,
  Ed.\hskip 1em plus 0.5em minus 0.4em\relax IGI Global, 2011, pp. 371--397.

\bibitem{Pearce2018}
M.~T. Pearce, ``Statistical learning and probabilistic prediction in music
  cognition: mechanisms of stylistic enculturation,'' \emph{Annals of the New
  York Academy of Sciences}, vol. 1423, no.~1, p. 378, 2018.

\bibitem{Allauzen1999}
C.~Allauzen, M.~Crochemore, and M.~Raffinot, ``Factor oracle: A new structure
  for pattern matching,'' in \emph{SOFSEMÕ99: Theory and Practice of
  Informatics}.\hskip 1em plus 0.5em minus 0.4em\relax Springer, 1999, pp.
  295--310.

\bibitem{Lefebvre2002}
A.~Lefebvre and T.~Lecroq, ``Compror: on-line lossless data compression with a
  factor oracle,'' \emph{Information Processing Letters}, vol.~83, no.~1, pp.
  1--6, 2002.

\bibitem{Dubnov2011}
S.~Dubnov, G.~Assayag, and A.~Cont, ``Audio oracle analysis of musical
  information rate,'' in \emph{Semantic Computing (ICSC), 2011 Fifth IEEE
  International Conference on}.\hskip 1em plus 0.5em minus 0.4em\relax IEEE,
  2011, pp. 567--571.

\bibitem{Luo2019}
Y.-J. Luo, K.~Agres, and D.~Herremans, ``Learning disentangled representations
  of timbre and pitch for musical instrument sounds using gaussian mixture
  variational autoencoders,'' 2019.

\bibitem{Hsu2017}
W.-N. Hsu, Y.~Zhang, and J.~Glass, ``Learning latent representations for speech
  generation and transformation,'' 2017.

\bibitem{Kingma2015}
D.~P. Kingma and J.~Ba, ``Adam: {A} method for stochastic optimization,'' in
  \emph{3rd International Conference on Learning Representations, {ICLR}},
  2015.

\bibitem{Wang2015}
C.-i. Wang and S.~Dubnov, ``Pattern discovery from audio recordings by variable
  markov oracle: A music information dynamics approach,'' in \emph{Acoustics,
  Speech, and Signal Processing (ICASSP), 2015 IEEE International Conference
  on}.\hskip 1em plus 0.5em minus 0.4em\relax IEEE, 2015.

\end{thebibliography}

\section{Appendix}
\appendix

\section{Audio VAE Architecture}
\label{appendix:a}
\begin{figure}[H]
    \centering
    \resizebox{0.5\linewidth}{!}{\includegraphics{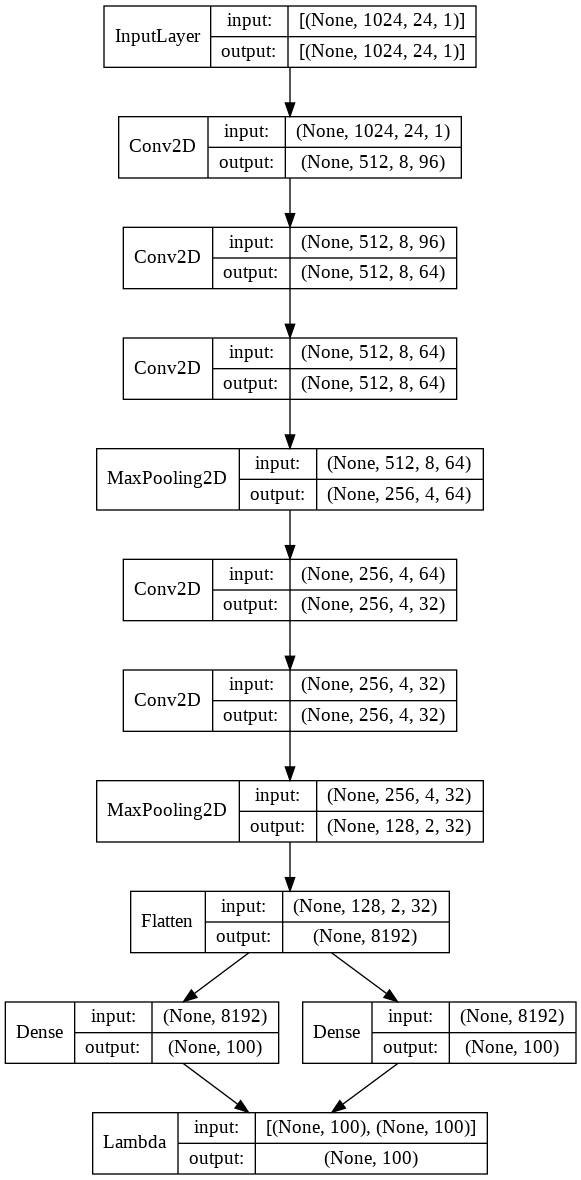}}
    \resizebox{0.4\linewidth}{!}{\includegraphics{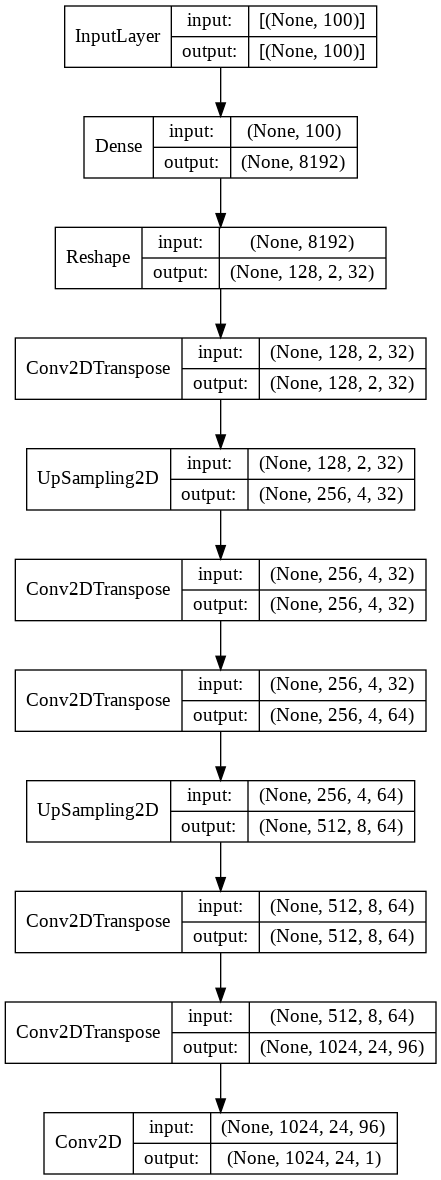}}
    \caption{Audio VAE Encoder (left) and Decoder (right)}
    \label{fig:vae_arch}
\end{figure}
As shown in Figure \ref{fig:vae_arch}, we use a simple symmetrical convolutional VAE. Both the encoder and decoder use 3 convolutional layers and max pooling. Fully connected layers are used to produce the latent distribution parameters, and to decode the latent representation into a downsampled image for decoding.

\section{East vs West Statistical Tests}
\label{appendix:b}

The ANOVA test shown in Table \ref{tab:anova_motif} shows significant difference in the mean pattern lengths between the select Eastern and Western music. Tables \ref{tab:first_hsd}-\ref{tab:last_has} show a followup Tukey HSD analysis on the pattern lengths for all pairs of songs in our selection.

\begin{table}[H]
    \centering
    \begin{tabular}{c|c c c}
Song Pair & Diff & p-adj & Reject\\\hline
Bach, Dongxiao & -8.30 & 0.001 & True \\
Bach, Shakuhachi & -10.10 & 0.001 & True \\
Bach, Telemann F. & -5.53 & 0.081 & False \\
Bach, Telemann R. & -5.55 & 0.031 & True \\
Dongxiao, Shakuhachi & -1.80 & 0.900 & False \\
Dongxiao, Telemann F. & 2.77 & 0.801 & False \\
Dongxiao, Telemann R. & 2.75 & 0.754 & False \\
Shakuhachi, Telemann F. & 4.57 & 0.552 & False \\
Shakuhachi, Telemann R. & 4.55 & 0.504 & False \\
Telemann F., Telemann R. & -0.02 & 0.900 & False \\
    \end{tabular}
    \caption{Tukey HSD Comparison of Means for $\text{VAE}_{all}$}
    \label{tab:first_hsd}
\end{table}

\begin{table}[H]
    \centering
    \begin{tabular}{c|c c c}
Song Pair & Diff & p-adj & Reject\\\hline
Bach, Dongxiao & -9.35 & 0.001 & True \\
Bach, Shakuhachi & -11.22 & 0.001 & True \\
Bach, Telemann F. & 0.01 & 0.900 & False \\
Bach, Telemann R. & -6.24 & 0.001 & True \\
Dongxiao, Shakuhachi & -1.87 & 0.900 & False \\
Dongxiao, Telemann F. & 9.36 & 0.001 & True \\
Dongxiao, Telemann R. & 3.11 & 0.481 & False \\
Shakuhachi, Telemann F. & 11.23 & 0.001 & True \\
Shakuhachi, Telemann R. & 4.98 & 0.205 & False \\
Telemann F., Telemann R. & -6.25 & 0.018 & True \\
    \end{tabular}
    \caption{Tukey HSD Comparison of Means for $\text{VAE}_{self}$}
    \label{tab:my_label}
\end{table}

\begin{table}[H]
    \centering
    \begin{tabular}{c|c c c}
Song Pair & Diff & p-adj & Reject\\\hline
Bach, Dongxiao & -3.23 & 0.006 & True \\
Bach, Shakuhachi & -5.04 & 0.001 & True \\
Bach, Telemann F. & 6.55 & 0.001 & True \\
Bach, Telemann R. & -1.09 & 0.762 & False \\
Dongxiao, Shakuhachi & -1.81 & 0.595 & False \\
Dongxiao, Telemann F. & 9.78 & 0.001 & True \\
Dongxiao, Telemann R. & 2.14 & 0.308 & False \\
Shakuhachi, Telemann F. & 11.59 & 0.001 & True \\
Shakuhachi, Telemann R. & 3.95 & 0.017 & True \\
Telemann F., Telemann R. & -7.64 & 0.001 & True \\
    \end{tabular}
    \caption{Tukey HSD Comparison of Means for $\text{VAE}_{exclude}$}
    \label{tab:my_label}
\end{table}

\begin{table}[H]
    \centering
    \begin{tabular}{c|c c c}
Song Pair & Diff & p-adj & Reject\\\hline
Bach, Dongxiao & -2.82 & 0.002 & True \\
Bach, Shakuhachi & -4.52 & 0.001 & True \\
Bach, Telemann F. & -3.70 & 0.001 & True \\
Bach, Telemann R. & 0.85 & 0.839 & False \\
Dongxiao, Shakuhachi & -1.70 & 0.430 & False \\
Dongxiao, Telemann F. & -0.89 & 0.894 & False \\
Dongxiao, Telemann R. & 3.66 & 0.002 & True \\
Shakuhachi, Telemann F. & 0.82 & 0.900 & False \\
Shakuhachi, Telemann R. & 5.37 & 0.001 & True \\
Telemann F., Telemann R. & 4.55 & 0.001 & True \\
    \end{tabular}
    \caption{Tukey HSD Comparison of Means for MFCC}
    \label{tab:my_label}
\end{table}

\begin{table}[H]
    \centering
    \begin{tabular}{c|c c c}
Song Pair & Diff & p-adj & Reject\\\hline
Bach, Dongxiao & -64.22 & 0.001 & True \\
Bach, Shakuhachi & -63.46 & 0.001 & True \\
Bach, Telemann F. & -40.82 & 0.001 & True \\
Bach, Telemann R. & -52.46 & 0.001 & True \\
Dongxiao, Shakuhachi & 0.76 & 0.900 & False \\
Dongxiao, Telemann F. & 23.40 & 0.001 & True \\
Dongxiao, Telemann R. & 11.77 & 0.123 & False \\
Shakuhachi, Telemann F. & 22.64 & 0.001 & True \\
Shakuhachi, Telemann R. & 11.01 & 0.318 & False \\
Telemann F., Telemann R. & -11.63 & 0.043 & True \\
    \end{tabular}
    \caption{Tukey HSD Comparison of Means for Chroma}
    \label{tab:last_has}
\end{table}

\end{document}